# Intervalley quantum interference and measurement of Berry phase in bilayer graphene

Yu Zhang[1, §], Ying Su[2, §], Lin He[1,†]

[1] Center for Advanced Quantum Studies, Department of Physics, Beijing Normal University, Beijing, 100875, People's Republic of China

[2] Theoretical Division, T-4 and CNLS, Los Alamos National Laboratory, Los Alamos, New Mexico 87545, USA

[§]These authors contributed equally to this work.
[†]Correspondence and requests for materials should be addressed to Lin He (e-mail: helin@bnu.edu.cn).

**Chiral quasiparticles in Bernal-stacked bilayer graphene have valley-contrasting Berry phases of ±2π. This nontrival topological structure, associated with the pseudospin winding along a closed Fermi surface, is responsible for various novel electronic properties, such as anti-Klein tunneling, unconventional quantum Hall effect, and valley Hall effect[1-6]. Here we show that the quantum interference due to intervalley scattering induced by atomic defects/impurities provides further insights into the topological nature of the bilayer graphene. The scattered chiral quasiparticles between distinct valleys with opposite chirality undergoes a rotation of pseudospin that results in the Friedel oscillation with wavefront dislocations. The number of dislocations reflects the information about pseudospin texture and hence can be used to measure the Berry phase[7]. As demonstrated both experimentally and theoretically, the Friedel oscillation, depending on the atomic defect/impurity at different sublattices, can exhibit $N$ = 4, 2, or 0 additional wavefronts, characterizing the 2π Berry phase of the bilayer graphene. Our results not only provide a comprehensive study of the intervalley quantum interference in bilayer graphene, but also shed lights on the pseudospin physics.**

Berry phase, a geometric phase accumulated by the wave function cycling around a closed path in momentum space, plays a very important role in determining the electronic dynamics and topological properties of materials[8-11]. In the low-energy band structures of graphene systems, the nonzero Berry phase is associated with the pseudospin winding[1-6,12-15]. In single-layer graphene (SLG), the pseudospin rotates by $2\pi$ along a closed Fermi surface that corresponds to the Berry phase $\gamma = \pi$, which is responsible for the Klein tunneling and half-integer quantum Hall effect[12-15]. In Bernal-stacked bilayer graphene (BLG), the pseudospin rotates by $4\pi$ along a closed Fermi surface and the Berry phase is $\gamma = 2\pi$. This leads to exotic electronic properties, such as the anti-Klein tunneling and integer quantum Hall effect with a peculiar absent zero Hall conductance plateau, in the BLG[1-6].

The direct measurement of the Berry phase is desirable to reveal the topological structure of materials and it is usually realized by a precise control of electromagnetic fields to drive quasiparticles to move adiabatically and coherently along a closed trajectory in momentum space[1,2,12-14]. Alternatively, probing the pseudospin winding also provides the information about the Berry phase of graphene systems. In the presence of time-reversal symmetry, the pseudospin texture has opposite chirality at distinct valleys that are time-reversal counterparts[7]. As a consequence, the scattered chiral quasiparticle between two distinct valleys undergoes a rotation of pseudospin to reverse its charity. The intervalley scattering thus gives rise to the quantum interference that encodes the information about the pseudospin winding and the Berry phase, and is manifested by the charge density modulation, i.e. the Friedel oscillation[7,16]. It has been shown in a recent experiment that the Berry phase of the SLG can be measured from the wavefront dislocations in the Friedel oscillation induced by an atomic impurity. The number of dislocations satisfies $2\pi N = 4\gamma$, from which the $N = 2$ additional wavefronts in the vicinity of an atomic impurity is a signature of the $\gamma = \pi$ Berry phase in the SLG[7]. This experiment provides an efficient method to measure the Berry phase without using electromagnetic fields.

In this work, we focus on the Bernal-stacked BLG whose topological structure is very different from the SLG. By using the scanning tunneling microscopy (STM), we measure the pseudospin winding and the Berry phase of the BLG from the wavefront dislocations in the Friedel oscillations around individual atomic defect/impurity. Unlike the SLG, the BLG exhibits $N$ = 4, 2, or 0 additional wavefronts for an atomic defect/impurity at different sublattices. However, according to $2\pi N = 4\gamma$, the Berry phase $\gamma = 2\pi$ of the BLG should result in $N$ = 4 additional wavefronts. To understand the observed peculiar phenomena, we simulate the intervalley quantum interference in the BLG with an atomic defect/impurity by the tight-binding model and continuum model. By projecting the charge density onto the two adjacent layers of the BLG, our theoretical simulation shows that the total number of dislocations in the two layers is $N$ = 4. However, the number of dislocations in the top layer can be $N$ = 4, 2, or 0, depending on the atomic defect/impurity at different sublattices of the BLG, as observed in our experiment.

In our experiments, we directly synthesize multilayer graphene on Ni foils by using a chemical vapor deposition (CVD) method[17-19] (see Methods and Supplementary Fig. S1 for more details). A high density of atomic impurities can be clearly observed via topographic STM images (Supplementary Fig. S2), which are attributed to the segregation mechanism for the graphene grown on the Ni foils and high carbon solubility of Ni. To identify the decoupled Bernal-stacked BLG regions, we carry out both the STM and magnetic-field-dependent scanning tunneling spectroscopy (STS) measurements. As observed in the Bernal-stacked BLG, the atomic-resolution STM images of the defect-free regions exhibit a triangular contrast due to the existence of sublattice asymmetry generated by the two adjacent graphene layers (Figs. 1a-1c)[20]. The high-magnetic-field STS spectra show well-defined Landau quantization of massive Dirac fermions (Supplementary Fig. S3), as observed previously in the Bernal-stacked BLG[20-22]. These experimental results demonstrate that the studied topmost BLG is efficiently decoupled from the underlying graphene sheets and behaves as the pristine Bernal-stacked BLG.

Figures 1a-1c show three representative STM images of the BLG with an atomic defect/impurity at different sublattices. The characteristic topographic fingerprint of the triangular $\sqrt{3} \times \sqrt{3}$ R30° interference patterns induced by the atomic defect/impurity can be clearly observed[18-19,23-26], as shown in Figs. 1a-1c. In the Bernal-stacked BLG, the B sublattices of the topmost graphene sheet are located exactly on top of the A′ sublattices of the underlying graphene layer. In the STM images, the visible atoms of the Bernal-stacked BLG are the A sublattice of the topmost layer[20,27,28]. Therefore, the located atomic sites of the defect/impurity can be unambiguously identified by overlying the lattice structures onto the STM images. The atomic structures can be further precisely determined via the apparent heights of the impurity-induced protrusions in the STM images[20] and orientations of the tripod shapes (white dotted outlines marked in Figs. 1a-1c) with respect to the directions of the nearest σ bonds of the single carbon defect (see Supplementary Fig. S4 and S5 for detail). According to our experiment, a single-carbon defect is located at the B′ (A) sublattice of the BLG in Fig. 1a (1c) and an impurity of hydrogen atom is chemisorbed on the B sublattice of the BLG in Fig. 1b, as marked by dashed circles and specified in Fig. 1d-1f, respectively.

The fast Fourier transform (FFT) of the STM images in Fig. 1a-1c are shown in Fig. 1g-1i, respectively. The bright spots with high intensity at the reciprocal lattice of the BLG are connected by the yellow dashed hexagon. At the center of the hexagon, the ring of high intensity with the radius of $2q_F$ (where $q_F$ is the Fermi wavevector) is induced by the intravalley scattering[29-31]. Such a feature is however suppressed in the SLG due to the π rotation of pseudospin and can be used to distinguish the BLG from the SLG[29-32]. The additional bright spots at the corners of Brillouin zone marked by green dashed hexagon are due to the intervalley scattering[29-33]. To show the intervalley scattering induced Friedel oscillations of charge density, we do the inverse FFT of the filtered bright spots enclosed by the white circles in Fig. 1g-1i. Figures 1j-1l show the corresponding FFT-filtered images for the cases of an atomic impurity located at the B′, B, and A sublattices, respectively. Here we exhibit only the charge density oscillation for the intervalley scattering in the direction specified by the white circles. The results for the intervalley scattering in other two directions are related by a $C_3$ rotation (see Supplementary Figs. S6-S8), which also show the same experimental features.

According to the experimental results in Figs. 1j-1l, there are apparent $N = 4, 2$, and $0$ additional wavefronts for the atomic defect/impurity at the B′, B, and A sublattices of the BLG, respectively.

To understand our experimental results, we first carry out the tight binding simulations to reproduce wavefront dislocations in the Friedel oscillation induced by intervalley scattering around atomic defects of the Bernal-stacked BLG (see Supplementary for details). We consider a BLG flake with $100 \times 100$ unit cells and periodic boundary conditions, and we locate the single carbon defect at the center. Because the STM measures mainly the charge density in the top layer, Fig. 2a-2c show the simulated charge densities in the top layer for the single carbon defect at the B′, B, and A sublattices of the BLG, as schematically shown in Figs. 1d-1f. Here we fix the Fermi energy $\omega = 30$ meV in the numerical simulation. In experiments, the Fermi energy is determined by the applied STM bias voltage. The FFT of the charge densities in Fig. 2a-2c are exhibited as insets. For the intervalley scattering in the direction specified by the white circles in the insets of Fig. 2a-2c, the filtered inverse FFT results in the charge density oscillations in Fig. 2d-2f. Here the numbers of additional wavefronts are $N = 4, 2$, and $0$ for the single-carbon defect at the B′, B, and A sublattices, respectively, and the results in other two directions, which are related by a $C_3$ rotation, exhibit the same features. Obviously, our theoretical results are well consistent with the experimental results, as shown in Fig. 1.

To gain further insights into the peculiar wavefront dislocations in the BLG, we study the atomic impurity induced intervalley quantum interference based on the low-energy continuum model of the BLG[6]

$$\mathcal{H}_{K_\xi + q} = -\frac{v_F^2 q^2}{t_\perp} \begin{pmatrix} 0 & e^{2\xi i \theta_q} \\ e^{-2\xi i \theta_q} & 0 \end{pmatrix} \quad (1)$$

which is expressed in the basis of $\{A, B'\}$. Here $\xi = \pm$ is the valley index, $v_F$ is the Fermi velocity, $t_\perp \approx 0.4$ eV is the nearest neighboring interlayer hopping[34], and $\theta_q$ is the polar angle of electrons with the momentum $q$ (see Fig. 3c). The eigenvalues of the Hamiltonian are $E^{\pm}_{K_\xi+q} = \pm v_F^2 q^2/t_\perp$ that yields the band structure in Fig. 3a. The corresponding eigenvectors are $|\psi^{\pm}_{K_\xi+q}\rangle = \frac{1}{\sqrt{2}}\begin{pmatrix}1 \\ \mp e^{-2\xi i\theta_q}\end{pmatrix}$ that defines the pesudospin

$$\langle\psi^{\pm}_{K_\xi+q}|\sigma|\psi^{\pm}_{K_\xi+q}\rangle = (\mp\cos 2\xi\theta_q, \pm\sin 2\xi\theta_q, 0) \qquad (2)$$

where $\sigma$ is the vector of Pauli matrices acting on the sublattice space. Figure 3b shows the pseudospin texture on the Fermi surfaces. The pseudospin rotates by $4\pi$ along a closed Fermi surface that yields the $W = 2$ winding number and the $\gamma = W\pi = 2\pi$ Berry phase in the BLG.

The intravalley and intervalley scattering processes in the BLG are sketched in Figs. 3c and 3d, respectively. Apparently, the intravalley scattering leaves the pseudospin unchanged (Fig. 3c), which can be negligible in the quantum interference. In contrast, the intervalley scattering rotates the pseudospin by $4\theta_q$ (Fig. 3d), which is twice lager than that in the SLG[32]. The schematic representation of the intervalley scattering process in real space is shown in Fig. 3e. At a given STM tip position, the amplitude of the Friedel oscillation is governed by the interference of the electronic waves in the incident and reflected processes. Here we parameterize the tip orientation relative to the impurity by the polar angle $\theta_r = \theta_q + \pi$, as shown in Fig. 3e. The locking of $\theta_r$ and $\theta_q$ indicates that circling the tip along a path enclosing the defect/impurity is equivalent to rotate $\theta_q$ along a closed Fermi surface. Because the pseudospin texture in Eq. (2) is originated from the phase difference between the two components of the eigenvector, the incident and reflected electronic waves have a phase difference of $4\theta_q$, same as the rotation angle of pseudospin in the intervalley backward scattering. Therefore, the phase shift accumulated over a closed scanning path enclosing the defect/impurity satisfies $\int_0^{2\pi} 4d\theta_q = 2\pi N = 4\gamma$. The defect/impurity at $r = 0$ can be regard as a phase singularity[35,36] and the total number of additional wavefronts due to the phase shift is $N = 4$ for the $\gamma = 2\pi$ Berry phase of the BLG.

However, the STM images, which mainly reflect the electronic distributions of the topmost graphene sheet in the BLG, are not necessary to show $N = 4$ additional wavefronts. In fact, the intervalley scattering induced modulation of charge densities projected onto the top and bottom graphene layers in the BLG are

$$\Delta\rho_t(\boldsymbol{r},\omega,\Delta\boldsymbol{K}) = \Delta\rho_A(r,\omega)\cos(\Delta\boldsymbol{K}\cdot\boldsymbol{r}+N_A\theta_r) + \Delta\rho_B(r,\omega)\cos(\Delta\boldsymbol{K}\cdot\boldsymbol{r}+N_B\theta_r), \quad (3)$$

$$\Delta\rho_b(\boldsymbol{r},\omega,\Delta\boldsymbol{K}) = \Delta\rho_{A'}(r,\omega)\cos(\Delta\boldsymbol{K}\cdot\boldsymbol{r}+N_{A'}\theta_r) + \Delta\rho_{B'}(r,\omega)\cos(\Delta\boldsymbol{K}\cdot\boldsymbol{r}+N_{B'}\theta_r), \quad (4)$$

where $\omega$ denotes the Fermi energy determined by the applied STM bias voltage and $\Delta\boldsymbol{K}$ is the momentum difference from a given $\boldsymbol{K}_+$ point to the nearest neighboring $\boldsymbol{K}_-$ point in the intervalley scattering process, as shown in Fig. 3d. $\Delta\rho_X(r,\omega)$ and $\cos(\Delta\boldsymbol{K}\cdot\boldsymbol{r}+N_X\theta_r)$ (X = A, B, A′, or B′) describe the modulation of charge density projected onto the X sublattice in the radial and transverse directions, respectively (see Supplementary for details). The charge density modulated by the trigonometric function changes signs for $|N_X|$ times when scanning around the defect/impurity with $\theta_r$ rotated by $2\pi$, thus resulting in the emergence of $N_X$ additional wavefronts in the Friedel oscillation at the X sublattice.

Now we focus on the low-energy limit, $\omega \ll t_\perp$, where $\Delta\rho_A \gg \Delta\rho_B$ and $\Delta\rho_{B'} \gg \Delta\rho_{A'}$ because the electronic states of the lowest two bands are mainly from the A and B′ sublattices. As a consequence, the wavefront dislocations in the top and bottom layers are expected to be determined by $|N_A|$ and $|N_{B'}|$ respectively. According to our calculation, $N_A = 4$ and $N_{B'} = 0$ for the single-carbon defect at the B′ sublattice, $N_A = 2$ and $N_{B'} = -2$ for the single-carbon defect at the B sublattice, $N_A = 0$ and $N_{B'} = -4$ for the single-carbon defect at the A sublattice (see Supplementary for details to calculate the $|N_A|$ and $|N_{B'}|$ for the single-carbon defect at the different sublattices). Indeed, there are $N$ = 4, 2, and 0 additional wavefronts in the top layer of the BLG for the defect/impurity at the B′, B, and A sublattices, as shown in Fig. 3f-3h. Such a result is consistent with the experimental and tight-binding results. Then we can provide a comprehensive understanding of our experimental results obtained in the Bernal-stacked BLG. The robust $N = |N_A| + |N_{B'}| = 4$ additional wavefronts are expected to be observed in the intervalley interference patterns in the BLG regardless of the atomic site of the defect/impurity. However, the defect/impurity can efficiently redistribute the numbers of additional wavefronts in the two adjacent graphene layers.

Therefore, we obtain $N$ = 4, 2, and 0 additional wavefronts in the top layer of the BLG in our experiment.

In summary, the intervalley scattering induced by atomic defects/impurities in Bernal-stacked BLG is systemically studied via the STM measurements. The quantum interference due to intervalley scattering is manifested by the wavefront dislocations in the Friedel oscillation that can be used to measure the Berry phase. In the BLG, the Berry phase $\gamma = 2\pi$ is characterized by the peculiar wavefront dislocations with $N$ = 4, 2 and 0 additional wavefronts for the atomic defect/impurity at the B′, B, and A sublattices in experiments. Our work provides a comprehensive measurement of the intervalley quantum interference and the Berry phase of the BLG. The method of measuring Berry phase from wavefront dislocations can be applied to multilayer graphene systems and other 2D materials.

## Methods

### Sample preparation.

A traditional low pressure CVD method was adopted to grow controllable layers of graphene on Ni(111) single crystal. The Ni(111) single crystal (5 mm ×5 mm width, 1 mm thickness) was first heated from room temperature to 900℃ in 40 min under an argon (Ar) flow of 100 SCCM (SCCM stands for Standard Cubic Centimeters per Minute) and hydrogen ($H_2$) flow of 100 SCCM, and keep this temperature and flow ratio for 20 min. Next $CH_4$ gas was introduced with a flow ratio of 20 SCCM, the growth time is ~15 min, and then cooled down to room temperature. Then samples are transferred into the ultrahigh vacuum condition for further characterizations.

### STM/STS measurements.

The STM system was an ultrahigh vacuum scanning probe microscope (USM-1300S and USM-1500S) from UNISOKU. All the STM and STS measurements were performed in the ultrahigh vacuum chamber (~$10^{-11}$ Torr) with constant-current scanning mode. The experiments were acquired at temperature of 4.2 K. The STM tips were obtained by chemical etching from a wire of Pt(80%) Ir(20%) alloys. Lateral dimensions observed in the STM images were calibrated using a standard graphene lattice and a Si (111)-(7×7) lattice and Ag (111) surface. The dI/dV measurements were taken with a standard lock-in technique by turning off the feedback circuit and using a 793-Hz 5mV a.c. modulation of the sample voltage.

## Acknowledgements

This work was supported by the National Natural Science Foundation of China (Grant Nos. 11974050, 11674029). L.H. also acknowledges support from the National Program for Support of Top-notch Young Professionals, support from "the Fundamental Research Funds for the Central Universities", and support from "Chang Jiang Scholars Program". Y.S. was supported by the U.S. Department of Energy through the Los Alamos National Laboratory LDRD program, and was supported by the Center for Non-linear Studies at LANL.

**Author contributions**

Y.Z. synthesized the samples, performed the STM experiments, and analyzed the data. Y.S. performed the theoretical calculations. L.H. conceived and provided advice on the experiment, analysis, and the theoretical calculation. Y.Z., Y.S., and L.H. wrote the paper. All authors participated in the data discussion.

**Competing financial interests**

The authors declare no competing financial interests.

# Figures

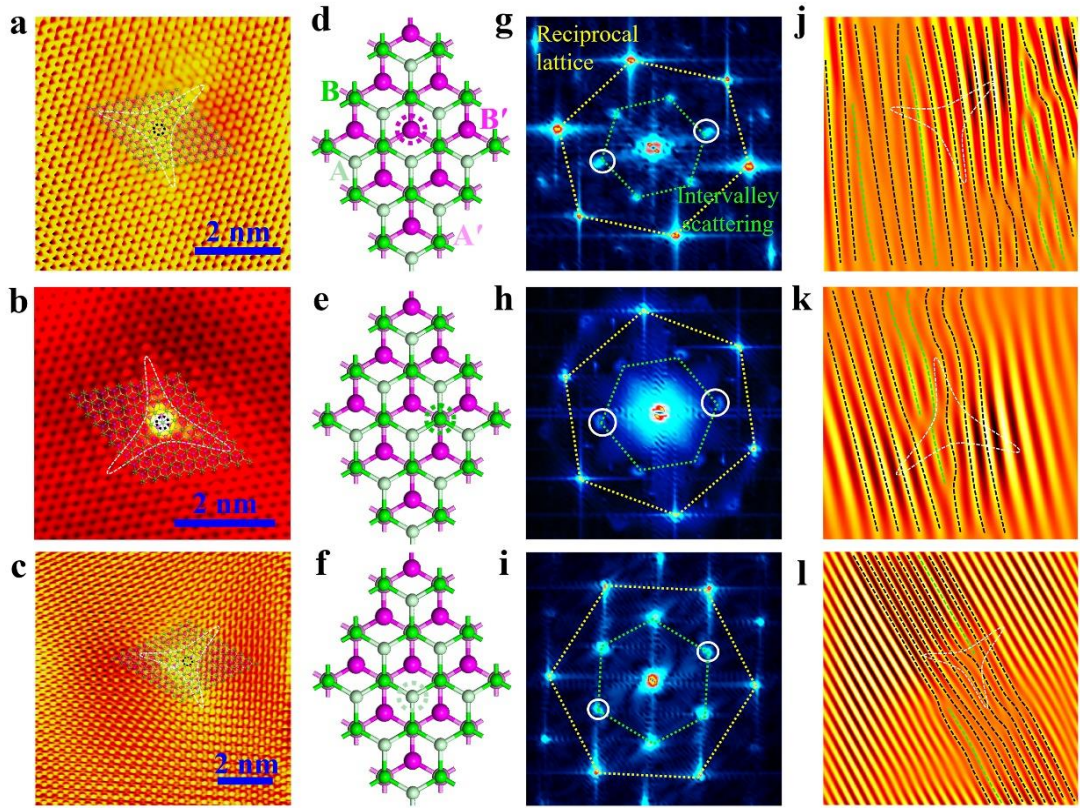

**Figure 1 | An individual single carbon defect/impurity induced wavefront dislocations in BLG. a-c,** The topography STM images of an individual single carbon defect/impurity in the decoupled BLG. The atomic structures of BLG are overlaid onto the STM images, illustrating the single carbon impurities are located on the B′ site (panel a), B site (panel b), and A site (panel c). **d-f,** Atomic structures of the BLG with the atomic defect/impurity at the B′, B, and A sublattices, respectively. The locations of the defect/impurity are marked by the dotted circles. **g-i,** FFT of the STM images in **a-c** respectively. The outer hexangular spots (corners of the yellow dotted line) and inner bright spots (corners of the green dotted line) correspond to the reciprocal lattice of graphene and the interference of the intervalley scattering, respectively. **j-k,** FFT-filtered images of panels **a-c** along the direction of intervalley scattering marked by white circles in panels **g-i**, which exhibit $N = 4$, 2, and 0 additional wavefronts (as marked by green dashed lines) respectively.

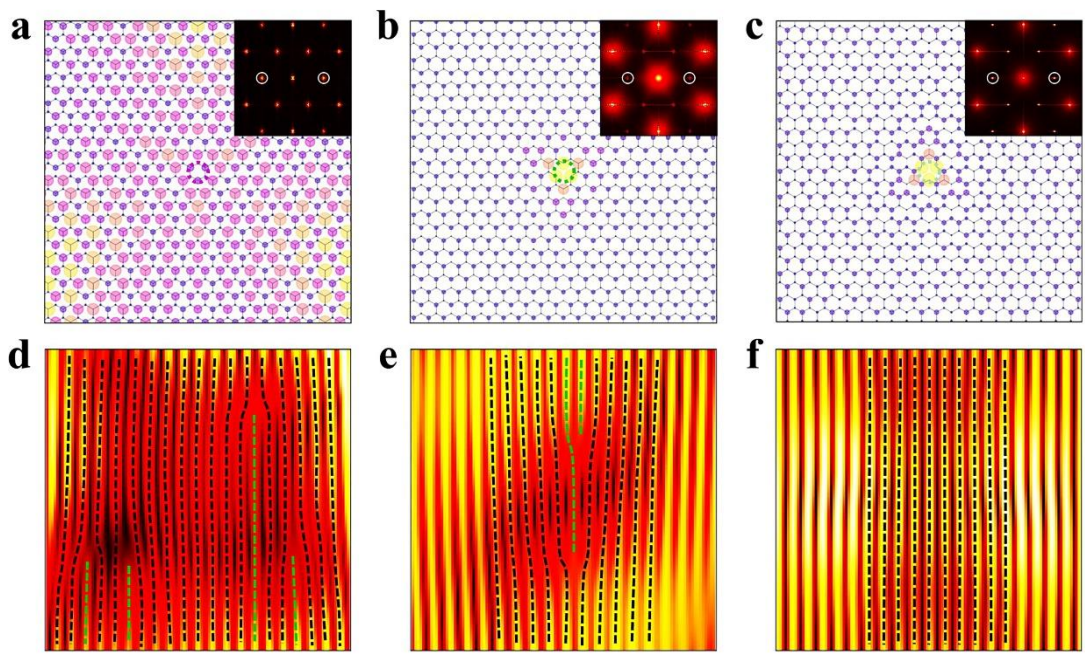

**Figure 2 | Tight binding calculations of charge density and wavefront dislocations induced by individual single carbon defect in BLG. a-c,** Simulated charge density in the top layer of the BLG with a single carbon impurity located at the B′, B, and A sublattices, respectively. Here the charge density is calculated at the energy of 30 meV and is represented by the spots whose size is proportional to the charge density. The FFT of the charge density is shown as insets. **d-f,** Charge density oscillations due to the intervalley quantum interference are obtained from the filtered FFT enclosed by white circles in a-c, respectively. For the single carbon impurity located at the three distinct sublattices in a-c, there are $N$ = 4, 2, and 0 additional wavefronts marked by green dashed lines.

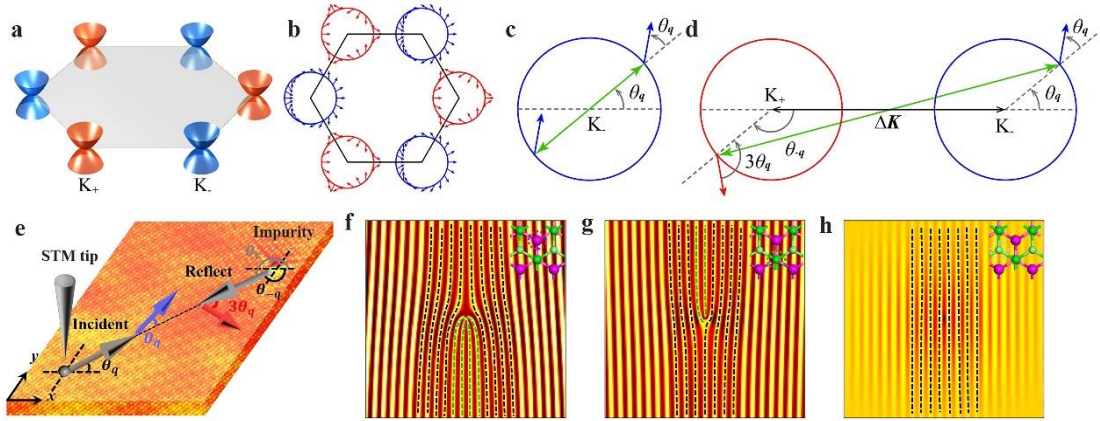

**Figure 3 | Low-energy continuum model calculations of wavefront dislocations from intervalley scattering in BLG. a,** Low-energy band structures around the Brillouin zone corners of the BLG. **b,** Pseudospin textures along the Fermi surfaces in the BLG. **c,** Schematic intravalley scattering in the BLG. **d,** Schematic intervalley scattering in the BLG. The pseudospin rotates by $4\theta_q$ in the intervalley scattering. **e,** The real space representation of the intervalley backward scattering in panel d. The red and blue arrows denote the pseudospins of reflected and incident quasiparticles from the $K_+$ and $K_-$ valleys, respectively. **f-h,** Charge density oscillations due to the intervalley scattering in panel d are calculated from the low-energy continuum model and projected onto the top layer of the BLG. For the single carbon impurity located at the B′, B, and A sublattices, as schematically shown in the insets, the charge density oscillations at the energy of 30 meV exhibit $N = 4$, 2, and 0 additional wavefronts, as marked by green dashed lines.